\documentclass[onecolumn,aps,12pt]{revtex4}
\usepackage{amsmath,graphicx,axodraw}
\usepackage{amsmath,graphicx} 
\def\,{\ifmmode\mskip\thinmuskip\else\leavevmode\thinspace\fi}

\newcommand{\dd}{\mbox{d}}
\newcommand\ba{\begin{eqnarray}}
\newcommand\ea{\end{eqnarray}}
\def\Li#1#2{{\mathrm{Li}}_{#1}\left(#2\right)}
\def\order#1{{\mathcal O}\left(#1\right)}
\def\la{\mathrel{\mathpalette\fun <}}

\def\fun#1#2{\lower3.6pt\vbox{\baselineskip0pt\lineskip.9pt
\ialign{$\mathsurround=0pt#1\hfil##\hfil$\crcr#2\crcr\sim\crcr}}}

\begin{document}

\title{Radiative muon (pion) pair production
in high energy electron-positron annihilation \\
(the case of small invariant pair mass)}

\author{A.B.~Arbuzov}
\affiliation{Joint Institute for Nuclear Research, 141980 Dubna,
Russia}
\author{E.~Barto\v{s}}
\altaffiliation{Department of Theoretical Physics, Comenius
University, 84248 Bratislava, Slovakia.}
\author{V.V.~Bytev}
\affiliation{Joint Institute for Nuclear Research, 141980 Dubna,
Russia}
\author{E.A.~Kuraev}
\affiliation{Joint Institute for Nuclear Research, 141980 Dubna,
Russia}
\author{Z.K.~Silagadze}
\affiliation{Budker Institute of Nuclear Physics, 630090 Novosibirsk, 
Russia}

\date{\today}

\begin{abstract}
The process of the muon (pion) pair production with small invariant mass
in the electron--positron high--energy annihilation, accompanied by
emission of hard photon at large angles, is considered. We find
that the Drell--Yan picture for differential cross section is valid
in the charge--even experimental set--up. Radiative corrections both
for electron block and for final state block are taken into account.
\end{abstract}

\maketitle

\section{Introduction}
\label{sect1}

Radiative return method, when the hard initial state radiation is used to 
reduce the invariant mass of a hadronic system produced in the high energy
electron-positron annihilation, provides an important tool to study various
hadronic cross-sections in a wide range of invariant masses without actually
changing the cms energy of the collider \cite{1,2,3,3a,4,5}. The very high 
luminosity of the modern meson factories makes the method competitive with 
the more conventional energy scan approach \cite{6,7}. Preliminary 
experimental studies both at KLOE \cite{8} and BABAR \cite{9} confirm the 
excellent potential of the radiative return method. It is not surprising, 
therefore, that the considerable efforts were devoted to elucidate the 
theoretical understanding of the radiative return process, especially for the
case of low energy pion pair production \cite{10,11,12,12a,12b,13,14}. 

The case, when the invariant mass of hadron system $\sqrt{s_1}$ is small 
compared to the center--of--mass total energy $\sqrt{s}=2\varepsilon$, 
represents a special interest. Such situation is realized, for example, in the
BABAR radiative return studies, where such interesting physical quantities
as form factors of the pion and the nucleon can be investigated.
The processes of radiative annihilation into muon and pion pairs, considered 
here, play a crucial role in such studies, both for the normalization purposes
and as one of the principal hadron production process at low energies.
Description of their differential cross sections with a rather high
level of accuracy (better than $0.5\%$ in the muon case) is the goal of our 
paper.

We specify the kinematics of the radiative muon (pion) pair creation process
\begin{eqnarray} \label{process}
&& e_-(p_-)+e_+(p_+)\to\mu_-(q_-)+\mu_+(q_+)+\gamma(k_1),
\end{eqnarray}
as follows:
\begin{eqnarray}
&& \qquad p_\pm^2=m^2,\quad q_\pm^2=M^2, \quad k_1^2=0,
\nonumber \\
&& \chi_\pm=2k_1\cdot p_\pm,\quad\chi_\pm'=2k_1\cdot q_\pm,\quad 
s=(p_-+p_+)^2,\quad s_1=(q_-+q_+)^2,
\nonumber \\
&& t=-2p_-\cdot q_-,\quad t_1=-2p_+\cdot q_+,\quad u=-2p_-\cdot q_+,\quad
u_1=-2p_+\cdot q_-,
\end{eqnarray}
where $m$ and $M$ are the electron and muon (pion) masses, respectively.
Throughout the paper we will suppose
\begin{eqnarray}
s\sim-t\sim-t_1\sim-u\sim-u_1\sim\chi_\pm\sim\chi_\pm'
\gg s_1> 4M^2\gg m^2.
\end{eqnarray}
Situation when $s\gg s_1 \gg M^2$ is also allowed.

\begin{figure}[htp]
\begin{picture}(300,160)(0,0)
\Vertex(100,70){3} \ArrowLine(0,70)(100,70)
\ArrowLine(200,70)(100,70) \Photon(100,70)(170,140){3}{10}
\Text(150,82)[]{$\vec{p}_+$} \Text(50,82)[]{$\vec{p}_-$}
\Text(130,118)[]{$\vec{k}$} \ArrowLine(100,70)(70,45)
\ArrowLine(100,70)(60,25) \Text(70,56)[]{$\vec{q}_-$}
\Text(90,40)[]{$\vec{q}_+$} \ArrowArcn(100,70)(30,180,45)
\Text(83,105)[]{$\theta$}
\end{picture}
\caption{Kinematical diagram for a radiative event with a small
invariant mass $s_1$.}
\label{kinem}
\end{figure}
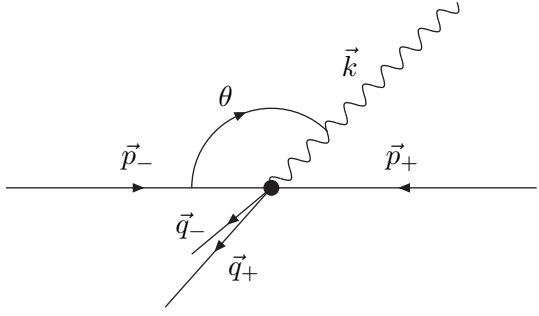

We will systematically omit the terms of the order of $M^2/s$ and $m^2/s_1$
compared with the leading ones. In $\order{\alpha}$
radiative corrections, we will drop also terms suppressed by the factor 
$s_1/s$. A kinematical diagram of the process
under consideration is drawn in Fig.~\ref{kinem}.
 
In this paper we will consider only the charge--even part of 
the differential cross section, which can be measured in an
experimental set--up blind to the charges of the created particles.
A detailed study of the charge--odd part of the radiative 
annihilation cross section in general kinematics will be 
presented elsewhere.

Our paper is organized as follows. The next Section is devoted to
the Born--level cross section. Radiative corrections to the final
and initial states are considered in Sect.~\ref{sect3}. This is followed
by concluding remarks. Some useful formulae are given in the appendix.

\section{The Born--Level Cross Section}
\label{sect2}

Within the Born approximation, the matrix element of the
initial state emission process has the form:
\begin{eqnarray}
M_B=\frac{(4\pi\alpha)^{3/2}}{s_1}\bar{v}(p_+)\biggl[
\gamma_\rho \frac{\hat{p}_- - \hat{k}_1+m}{-2p_-k_1}\hat{e}
+ \hat{e}\frac{-\hat{p}_++\hat{k_1}+m}{-2p_+k_1}\gamma_\rho
\biggr] u(p_-) J^\rho
\end{eqnarray}
with
\begin{eqnarray}
J^\rho=\bar{u}(q_-)\gamma^\rho u(q_+)
\end{eqnarray}
for the muon pair production, and
\begin{eqnarray}
J^\rho_\pi=(q_- - q_+)^\rho F_\pi^{str}(s_1)
\end{eqnarray}
for the case of charged pions, $F_\pi^{str}(s_1)$ being the pion
strong interaction form factor.

The corresponding contribution to the cross section is
\begin{eqnarray} \label{born1}
&& \frac{\dd\sigma_B^{j}}{\dd\Gamma}=\frac{\alpha^3}{8\pi^2ss_1^2}R^j,
\qquad R^{j}=B^{\rho\sigma} i^{(0j)}_{\rho\sigma}, \qquad
i^{(0j)}_{\rho\sigma} = \sum_{pol} J_{\rho}^j({J_{\sigma}}^j)^\star,\quad 
j=\mu, \pi,
\nonumber \\
&& B_{\rho\sigma}= B_gg_{\rho\sigma} + B_{11}(p_{-}p_{-})_{\rho\sigma} +
B_{22}(p_{+}p_{+})_{\rho\sigma},
\nonumber \\
&& B_g=-\frac{(s_1+\chi_+)^2+(s_1+\chi_-)^2}{\chi_+\chi_-}, \quad
B_{11}=-\frac{4s_1}{\chi_+\chi_-},\quad
B_{22}=-\frac{4s_1}{\chi_+\chi_-},
\end{eqnarray}
where we have used the short hand notations $(qq)_{\rho\sigma} = 
q_{\rho}q_{\sigma}$,
$(pq)_{\rho\sigma} = p_{\rho}q_{\sigma} + q_{\rho}p_{\sigma}$. For the muon 
final state 
\ba 
i_{\rho\sigma}^{(0\mu)} =
4\biggl[(q_+q_-)_{\rho\sigma} - g_{\rho\sigma}\frac{s_1}{2}\biggr]. 
\ea 
For the case of pions,
\ba 
i_{\rho\sigma}^{(0\pi)} = |F_\pi^{str}(s_1)|^2(q_--q_+)_{\rho}
(q_--q_+)_{\sigma}. 
\ea

Note that the Born-level cross section for the $e^+e^-\to\mu^+\mu^-\gamma$
process was calculated in \cite{15,16} (see also \cite{17}).

%
%

The phase space volume of the final
particles is
\begin{eqnarray}
\dd\Gamma=\frac{\dd^3q_-}{\varepsilon_-}\frac{\dd^3q_+}{\varepsilon_+}
\frac{\dd^3k_1}{\omega_1}\delta^4(p_++p_--q_+-q_--k_1).
\end{eqnarray}
For the case of small invariant mass of the created pair $s_1\ll s$, it
can be rewritten as (see fig.\ref{kinem}):
\begin{gather}
\dd\Gamma=\pi^2\dd x_-\dd c \dd s_1,
\end{gather}
(note that $s_1$ is small due to $c\to 1$, but the energy of muon pair is
large: $sx_\pm^2\gg 4M^2$)
and approximately
\begin{eqnarray}
&& x_\pm=\frac{\varepsilon_\pm}{\varepsilon}, \quad x_++x_-=1,\quad
\omega_1=\frac{1}{2}\sqrt{s},\quad
\chi_\pm=\frac{s(1\mp c )}{2}, \quad \chi_\pm'=sx_\pm,
\nonumber \\ \nonumber
&& t=\frac{-sx_-(1-c)}{2} ,\quad
t_1=\frac{-sx_+(1+c)}{2},\quad u=\frac{-sx_+(1-c)}{2},\quad
u_1=\frac{-sx_-(1+c)}{2}\, , \\ \nonumber
&&c=\cos(\widehat{\vec{p}_-\vec{q}}_-)=\cos\theta.
\end{eqnarray}
We will assume that the emission angle of the hard photon lies
outside the narrow cones around the beam axis: $\theta_0 <\theta_1 <
\pi-\theta_0 $, with $\theta_0\ll1$, $\theta_0\varepsilon\gg
M$.

When the initial state radiation dominates the Born cross-section 
take a rather simple forms:
\begin{eqnarray} \label{born0}
&& \dd\sigma_B^{(\mu)}(p_-,p_+;k_1,q_-,q_+)=
\frac{\alpha^3(1+c^2)}{ss_1(1-c^2)}
\biggl[2\sigma+1-2x_-x_+\biggr]\dd x_-\dd c \dd s_1,
\\ \nonumber
&& \dd\sigma_B^{(\pi)}(p_-,p_+;k_1,q_-,q_+)=
\frac{\alpha^3(1+c^2)}{ss_1(1-c^2)}\,|F_\pi^{str} (s_1)|^2\,
\biggl[-\sigma+x_-x_+\biggr]\dd x_-\dd c \dd s_1,
\\ \nonumber
&& \frac{1}{2}(1-\beta)<x_-<\frac{1}{2}(1+\beta),\quad
\beta=\sqrt{1-\frac{4M^2}{s_1}}, \quad \sigma=\frac{M^2}{s_1}.
\end{eqnarray}
Here $\beta$ is the velocity of the pair component in the center--of--mass
reference frame of the pair.

\section{Radiative corrections}
\label{sect3}

Radiative corrections (RC) can be separated
into three gauge--invariant parts.
They can be taken into
account by the formal replacement (see(\ref{born1})):
\ba \label{zzz}
\frac{R^j}{s_1^2}\longrightarrow
\frac{K^{\rho\sigma}J_{\rho\sigma}^j}{s_1^2|1-\Pi(s_1)|^2}
\ea
where $\Pi(s_1)$ describes the vacuum polarization of the
virtual photon (see Appendix); $K^{\rho\sigma}$ is the initial--state
emission Compton tensor with RC taken into account; 
$J_{\rho\sigma}^j$ is the final state current tensor with 
$\order{\alpha}$ RC. 

First we consider the explicit formulae for RC due to 
virtual, soft, and hard collinear final state emission. 
As concerning RC to the initial state 
for the charge--blind experimental set--up considered here,  
we will use the explicit expression for the Compton
tensor with heavy photon $K^{\rho\sigma}$
calculated in the paper \cite{19} for the scattering channel and apply the 
crossing transformation (see also \cite{12a}). Possible contribution due to 
emission of an additional real photon from the initial state will be
taken into account too.
In conclusion we will give
the explicit formulae for the cross section, consider separately the
kinematics of the collinear emission, and estimate the contribution of
higher orders of perturbation theory (PT).

\subsection{Corrections to the final state}

The third  part is related to the lowest order RC to the muon (pion)
current \ba J_{\rho\sigma}= i^{(v)}_{\rho\sigma} + i^{(s)}_{\rho\sigma} +
i^{(h)}_{\rho\sigma}. \ea The virtual photon contribution
$i^{(v)}_{\rho\sigma} $ takes into account the Dirac and Pauli form
factors of the muon current
\begin{eqnarray}
J^{(v\mu)}_\rho=\bar{u}(q_-)[\gamma_\rho F_1(s_1)+
\frac{\hat{q}\gamma_\rho-\gamma_\rho\hat{q}}{4M}F_2(s_1)]v(q_+),
\quad q = q_+ + q_-, \quad s_1=q^2.
\end{eqnarray}
We have
\begin{eqnarray}
B^{\rho\sigma} i^{(v\mu)}_{\rho\sigma} = B_g \sum_{pol}|J^{(v\mu)}_{\rho}|^2 +
B_{11}\biggl[\sum_{pol}|p_-\cdot J^{(v\mu)}|^2 + \sum_{pol}|p_+\cdot J^{(v\mu)}|^2\biggr].
\end{eqnarray}
Here $\Sigma$ means a sum over the muon spin states and
\begin{eqnarray} \label{ffa}
&& \sum_{pol}|J^{(v\mu)}_\rho|^2=\frac{\alpha}{\pi}\biggl[-8(s_1+2M^2)f_1^{(\mu)}
- 12s_1f_2^{(\mu)}\biggr],
\nonumber \\
&& \sum_{pol} |J^{(v\mu)}\cdot p_\pm|^2=\frac{\alpha}{\pi}s^2(1\pm c)^2
(x_+x_-f_1^{(\mu)}+ \frac{1}{4}f_2^{(\mu)}).
\end{eqnarray}
 For the pion final state we have \ba &&
B^{\rho\sigma}i_{\rho\sigma}^{(v\pi)} =2\frac{\alpha}{\pi} B^{\rho\sigma} 
i_{\rho\sigma}^{(0\pi)} f^{QED}_{\pi},
\nonumber \\
&& B^{\rho\sigma} i^{(0\pi)}_{\rho\sigma} = 2\frac{\alpha}{\pi} |F_\pi^{str} (s_1)|^2
\biggl[(4M^2-s_1)B_g + \frac{1}{8}s^2B_{11}(x_+-x_-)^2(1+c^2)\biggr] f^{QED}_{\pi}.
\ea
The explicit expression for the $f_{1,2}^\mu $, $f^{QED}_{\pi}$
form factors of pion and muon are given in the Appendix.

The soft photon correction to the final state currents reads \ba
&& i_{\rho\sigma}^{(s\pi)}
=\frac{\alpha}{\pi}\Delta_{1'2'}i_{\rho\sigma}^{(0\pi)},
\quad\quad\quad i_{\rho\sigma}^{(s\mu)}=
\frac{\alpha}{\pi}\Delta_{1'2'}i_{\rho\sigma}^{(0\mu)},
\nonumber \\
&& \Delta_{1'2'} = - \frac{1}{4\pi}\int\frac{\dd^3k}{\omega}
\biggl(\frac{q_+}{q_+k} - \frac{q_-}{q_-k}\biggr)^2
\bigg|_{\omega\leq\Delta\varepsilon} =
(\frac{1+\beta^2}{2\beta}\ln\frac{1+\beta}{1-\beta}-1)
\ln\frac{(\Delta\varepsilon)^2 M^2}{\varepsilon^2 x_+x_-\lambda^2}
\\ \nonumber && \qquad
+ \frac{1+\beta^2}{2\beta}\biggl[ - g -
\frac{1}{2}\ln^2\frac{1+\beta}{1-\beta} -
\ln\frac{1+\beta}{1-\beta}\ln\frac{1-\beta^2}{4} - \frac{\pi^2}{6}
- 2\Li{2}{\frac{\beta-1}{\beta+1}} \biggr],
\\ \nonumber &&
g=2\beta\int\limits_0^1\frac{\dd t}{1-\beta^2t^2}
\ln\biggl(1+\frac{1-t^2}{4}\,\frac{(x_+ - x_-)^2}{x_+x_-}\biggr) =
\ln\biggl(\frac{1+\beta}{1-\beta}\biggr)\ln(1 + z -z/\beta^2)
\\ \nonumber && \qquad
+ \Li{2}{\frac{1-\beta}{1+\beta/r}} +
\Li{2}{\frac{1-\beta}{1-\beta/r}} -
\Li{2}{\frac{1+\beta}{1-\beta/r}}
- \Li{2}{\frac{1+\beta}{1+\beta/r}},
\\ \nonumber &&
\beta =\sqrt{1-\frac{4M^2}{s_1}},\qquad z =
\frac{1}{4}\biggl(\sqrt{\frac{x_+}{x_-}} -
\sqrt{\frac{x_-}{x_+}}\biggr)^2, \qquad r =
\left|x_+-x_-\right|.
\ea
This formulae provides the generalization of known expression 
(see (25,26) in \cite{viol}) for the case of small invariant mass $4M^2\sim \sqrt{s_1}\ll \varepsilon_\pm$

The contribution of an additional hard photon emission (with momentum $k_2$)
by the muon block, provided $\tilde{s}_{1}=(q_++q_-+k_2)^2\sim s_1\ll s$,
can be found by the expression
\begin{eqnarray}
B^{\rho\sigma}i_{\rho\sigma}^{(h\mu)} = \frac{\alpha}{4\pi^2}
\int\frac{\dd^3k_2}{\omega_2}B^{\rho\sigma}\sum J_\rho^{(\gamma)}
(J_{\sigma}^{(\gamma)})^*
\bigg|_{\omega_2 \geq \Delta\varepsilon},
\end{eqnarray}
with
\begin{eqnarray}
&& \sum|J^{(\gamma)}_{\rho}|^2=4Q^2(s_1+2k_2\cdot q_-+2k_2\cdot q_++2M^2)-
8\frac{(k_2\cdot q_-)^2+(k_2\cdot q_+)^2}{(k_2\cdot q_-)(k_2\cdot q_+)},
\nonumber \\
&& Q=\frac{q_{-}}{q_-\cdot k_2}-\frac{q_{+}}{q_+\cdot k_2},
 \nonumber
\end{eqnarray}
and
\begin{eqnarray}
&& \sum|J^{(\gamma)}\cdot p_\pm|^2 = - 8Q^2(q_-\cdot p_\pm)(q_+\cdot p_\pm) 
+ 8(p_\pm\cdot k_2)\biggl(Q\cdot q_+\frac{p_\pm\cdot q_-}{q_+\cdot
k_2}-Q\cdot q_- \frac{p_\pm\cdot q_+}{q_-\cdot k_2}\biggr)
\nonumber \\ && \hspace*{-3mm}
+  8(p_\pm\cdot k_2)\biggl(\frac{p_\pm\cdot q_-}
{q_+\cdot k_2}+ \frac{p_\pm\cdot
q_+}{q_-\cdot k_2}\biggr) + 8(p_\pm\cdot Q)(p_\pm\cdot q_+-p_\pm\cdot
q_-)-8\frac{(k_2\cdot p_\pm)^2M^2}{(k_2\cdot q_+)(k_2\cdot q_-)}\; .
\end{eqnarray}

For the case of charged pion pair production,
the radiative current tensor has the form
\begin{gather}
i_{\rho\sigma}^{(h\pi)} = -\frac{\alpha}{4\pi^2}\int
|F_\pi^{str}(\tilde{s}_{1})|^2 \frac{\dd^3k_2}{\omega_2}
\biggl[\frac{M^2}{\chi_{2-}^2}(Q_1Q_1)_{\rho\sigma} +
\frac{M^2}{\chi_{2+}^2}(Q_2Q_2)_{\rho\sigma}-
\frac{q_+q_-}{\chi_{2+}\chi_{2-}}(Q_1Q_2)_{\rho\sigma}
\nonumber\\
+g_{\rho\sigma}- \frac{1}{\chi_{2-}}(Q_1q_-)_{\rho\sigma}
+\frac{1}{\chi_{2+}}(Q_2q_+)_{\rho\sigma}\biggr]
\bigg|_{\omega_2>\Delta\varepsilon},
\\ \nonumber
Q_1=q_--q_++k_2,\qquad  Q_2=q_--q_+-k_2, \qquad \chi_{2\pm} =
2k_2\cdot q_{\pm}.
\end{gather}
One can check that the Bose symmetry and the
gauge invariance condition is valid for the pionic current tensor.
Namely it is invariant with regard to the permutation of the pion momenta
and turns to zero after conversion with 4-vector $q$.

The sum of soft and hard photon corrections to the final current does not 
depend on $\Delta\varepsilon/\varepsilon$.

\subsection{Corrections to the initial state}

Let us now consider the Compton tensor with RC, which describe
virtual corrections to the initial state.
In our kinematical region it will be
convenient to rewrite the tensor explicitly extracting large
logarithms. We will distinguish two kinds of large logarithms:
\begin{gather}
l_s=\ln\frac{s}{m^2},\quad l_1=\ln\frac{s}{s_1}.
\end{gather}
We rewrite the Compton tensor \cite{19} in the form:
\begin{gather}
\label{comp}
K_{\rho\sigma}=(1+\frac{\alpha}{2\pi}\rho)B_{\rho\sigma}+
\frac{\alpha}{2\pi}\bigl[\tau_gg_{\rho\sigma}
+\tau_{11}(p_{-}p_{-})_{\rho\sigma}+\tau_{22}(p_{+}p_{+})_{\rho\sigma}-
\frac{1}{2}\tau_{12}(p_{-}p_{+})_{\rho\sigma}
\biggr], \\ \nonumber
\rho=-4\ln\frac{m}{\lambda}(l_s-1)-l_s^2+3l_s-3l_1+\frac{4}{3}\pi^2-
\frac{9}{2},
\end{gather}
with $\tau_i=a_il_1+b_i$ and \ba a_{11} &=&
-\frac{2s_1}{\chi_+\chi_-}\biggl[\frac{2b^2}{\chi_+\chi_-}+\frac{4s}{a}+
\frac{4(s^2+b\chi_-)}{a^2}-\frac{b^2(2c-\chi_-)}{c^2\chi_-}-\frac{2s+
\chi_+}{\chi_+}\biggr],
\\
b_{11} &=& \frac{2}{\chi_+\chi_-}\biggl[-s_1(1+\frac{s^2}{\chi_+^2})G_-
-s_1\biggl(2+\frac{b^2}{\chi_-^2}\biggr)G_+
-\frac{s_1b^2(2c-\chi_-)}{c^2\chi_-}\ln\frac{s}{\chi_+}
\nonumber \\
&-& \frac{s_1}{\chi_+}(2s+\chi_+)\ln\frac{s}{\chi_-}
-\frac{4(s^2+b\chi_-)}{a}-4s-2s_1-\chi_--\frac{b^2}{c} \biggr],
\ea \ba a_{12} &=&-
\frac{2s_1}{\chi_+\chi_-}\biggl[-\frac{4ss_1}{\chi_+\chi_-}
+\frac{8(\chi_+\chi_--s^2)}{a^2}
-\frac{4s}{a}+4ss_1\biggl(\frac{1}{c\chi_-}+\frac{1}{b\chi_+}\biggr)
\nonumber \\
&+& (2ss_1+4\chi_+\chi_-)\biggl(\frac{1}{c^2}+\frac{1}{b^2}\biggr) \biggr],
\\
b_{12}&=&\frac{2}{\chi_+\chi_-}\biggl[\frac{2s_1}{\chi_+^2}(sc-\chi_-\chi_+)G_-
+\frac{2s_1}{\chi_-^2}(sb-\chi_-\chi_+)G_+
\nonumber \\
&+& s_1\biggl(\frac{2ss_1+4\chi_-\chi_+}{c^2}+\frac{4ss_1}{c\chi_-}\biggr)
\ln\frac{s}{\chi_+}
+s_1\biggl(\frac{2ss_1+4\chi_-\chi_+}{b^2}+\frac{4ss_1}{b\chi_+}\biggr)
\ln\frac{s}{\chi_-} \nonumber \\
&+&
\frac{8(s^2-\chi_+\chi_-)}{a}-2s\biggl(\frac{\chi_+}{c}+
\frac{\chi_-}{b}\biggr)+2s_1+10s
\biggr], \ea \ba a_g &=&
-2s\biggl(\frac{s_1}{\chi_+\chi_-}-\frac{2}{a}\biggr)
+\frac{c}{\chi_+}\biggl(\frac{3s}{b}-1\biggr)
+\frac{b}{\chi_-}\biggl(\frac{3s}{c}-1\biggr),
\\
b_g &=& -\frac{1}{\chi_+}(\frac{ss_1}{\chi_+}+\frac{2sb+\chi_+^2}{\chi_-})G_-
-\frac{1}{\chi_-}(\frac{ss_1}{\chi_-}+\frac{2sc+\chi_-^2}{\chi_+})G_+
\nonumber \\
&-& \frac{c}{\chi_+}\biggl(\frac{3s}{b}-1\biggr)\ln\frac{s}{\chi_-}
-\frac{b}{\chi_-}\biggl(\frac{3s}{c}-1\biggr)\ln\frac{s}{\chi_+}
+\frac{2s^2-\chi_+^2-\chi_-^2}{2\chi_+\chi_-},
\\ \nonumber
a&=&-(\chi_++\chi_-),\quad b=s_1+\chi_-,\quad c=s_1+\chi_+,
\\ \nonumber
G_-&=&-\ln^2\frac{\chi_-}{s}+\frac{\pi^2}{3}-2\Li{2}{1-\frac{s_1}{s}}
+2\Li{2}{-\frac{s_1}{\chi_-}}+2\ln\frac{s_1}{\chi_-}\ln\biggl(1+\frac{s_1}{\chi_-}\biggr),
\\ \nonumber
G_+ &=& -\ln^2\frac{\chi_+}{s} +
\frac{\pi^2}{3}-2\Li{2}{1-\frac{s_1}{s}}+2\Li{2}{-\frac{s_1}{\chi_+}}
+2\ln\frac{s_1}{\chi_+}\ln\biggl(1+\frac{s_1}{\chi_+}\biggr),
\\ \nonumber
&&\tau_{22}(\chi_-,\chi_+)=\tau_{11}(\chi_+,\chi_-).
 \ea

The infrared singularity (the presence of the {\em\ photon mass} $\lambda$ in
$\rho$) is compensated by taking into account soft
photon emission from the initial particles:
\begin{gather}
\dd\sigma^{\mathrm{soft}}=\dd\sigma_0\frac{\alpha}{\pi}\Delta_{12},\\
\nonumber \Delta_{12}=-\frac{1}{4\pi}\int\frac{\dd^3k}{\omega}
\biggl(\frac{p_+}{p_+k} - \frac{p_-}{p_-k}\biggr)^2
\bigg|_{\omega\leq\Delta\varepsilon}=2(l_s-1)
\ln\frac{m\Delta\varepsilon}{\lambda\varepsilon}
+\frac{1}{2}l_s^2-\frac{\pi^2}{3}.
\end{gather}
As a result, the quantity $\rho$ in formula (\ref{comp}) will change to
\begin{gather}
\rho\to\rho_\Delta=(4\ln\frac{\Delta\varepsilon}{\varepsilon}+3)(l_s-1)-3l_1
+ \frac{2\pi^2}{3}-\frac{3}{2}\, .
\end{gather}

Cross section of two hard photon emission for the case when
one of them is emitted collinearly to the incoming electron or positron
can be obtained by means of the quasi--real electron method~\cite{Baier}:
\begin{eqnarray}
\frac{\dd\sigma_{\gamma\gamma , \,coll}^{j}}{\dd x_-\dd c\;\dd s_1} &=&
\dd W_{p_-}(k_3)\frac{\dd\tilde\sigma_B^j (p_-(1-x_3),p_+;k_1,q_+,q_-)}
{\dd x_-\dd c\;\dd s_1}
\nonumber \\ 
&+& \dd W_{p_+}(k_3)\frac{\dd\tilde\sigma_B^j(p_-,p_+(1-x_3);k_1,q_+,q_-)}
{\dd x_-\dd c\;\dd s_1},
\end{eqnarray}
with
\begin{eqnarray}
\dd W_p(k_3)=\frac{\alpha}{\pi}[(1-x_3+\frac{x_3^2}{2})
\ln\frac{(\varepsilon\theta_0)^2}{m^2}-(1-x_3)]\frac{\dd x_3}{x_3},
\qquad x_3=\frac{\omega_3}{\varepsilon},\quad
x_3>\frac{\Delta\varepsilon}{\varepsilon}.
\end{eqnarray}
Here we suppose that the polar angle $\theta_3$ between the
directions of the additional collinear photon and the beam axis  does not 
exceed some small value $\theta_0\ll 1$, $\ \varepsilon\theta_0\gg m$.

The {\em boosted} differential cross section 
$\dd\tilde\sigma_B^j(p_-x,p_+y;k_1,q_+,q_-)$
with reduced momenta of the incoming particles reads 
(compare with Eq.~(\ref{born0}))
\begin{eqnarray} \label{boosted}
&& \frac{\dd\tilde\sigma_B^\mu(p_+x_2,p_-x_1;k_1,q_+,q_-)}
{\dd x_-\dd c\dd s_1}=
\frac{\alpha^3(1+2\sigma-2\nu_-(1-\nu_-))(x_1^2(1-c)^2+x_2^2(1+c)^2)}
{s_1sx_1^2x_2^2(1-c^2)(x_1+x_2+c(x_2-x_1))},
\nonumber \\
&& \frac{\dd\tilde\sigma_B^\pi(p_+x_2,p_-x_1;k_1,q_+,q_-)}
{\dd x_-\dd c\dd s_1}=
\frac{\alpha^3(\nu_-(1-\nu_-)-\sigma)(x_1^2(1-c)^2+x_2^2(1+c)^2)}
{s_1sx_1^2x_2^2(1-c^2)(x_1+x_2+c(x_2-x_1))},
\nonumber \\
&& \nu_-=\frac{x_-}{y_2},\qquad
y_2=\frac{2x_1x_2}{x_1+x_2+c(x_2-x_1)}.
\end{eqnarray}

In a certain experimental situation, an estimate of the contribution of
the additional hard photon emission outside the narrow cones around the beam 
axes is needed. It can be estimated by
\ba
&& \frac{\dd\sigma_{\gamma\gamma , noncoll}^{j}}{\dd x_-\dd c\;\dd s_1}
= \frac{\alpha}{4\pi^2} \int\frac{\dd^3k_3}{\omega_3}
\biggl[\frac{\varepsilon^2+(\varepsilon-\omega_3)^2}
{\varepsilon\omega_3}\biggr] \biggl\{
\frac{1}{k_3\cdot p_-}\frac{\dd\sigma_B^j(p_-(1-x_3),p_+;k_1,q_+,q_-)}
{\dd x_-\dd c\;\dd s_1} \nonumber
\\ && \qquad
+ \frac{1}{k_3\cdot p_+}\frac{\dd\sigma_B^j(p_-,p_+(1-x_3);k_1,q_+,q_-)}
{\dd x_-\dd c\;\dd s_1} \biggr\}
\bigg|_{\theta_3\geq\theta_0,\ \
\Delta\varepsilon<\omega_3<\omega_1}, \qquad x_3 = 
\frac{\omega_3}{\varepsilon}\, .
\ea
It is a simplified expression for the two--photon initial state
emission cross section. Deviation, for the case of a large angle
emission, of our estimate from the exact result is small. 
It does not depend on $s$ and slightly depends on $\theta_0$.
For $\theta_0 \sim 10^{-2}$ we have
\begin{eqnarray}
\frac{\pi}{\alpha}\bigg|\frac{\int(\dd\sigma_{\gamma\gamma,\,noncoll}^j-
\dd\sigma_{\gamma\gamma,\,
noncoll \, exact}^j)} {\int\dd\sigma^j_B}\bigg| \la 10^{-1}.
\end{eqnarray}

\subsection{Master formula}

By summing up all contributions for the charge--even part, we can put the
cross section of the radiative production in the form:
\ba \label{all}
&& \frac{\dd\sigma^{j}(p_+,p_-;k_1,q_+,q_-)}{\dd x_-\dd c\;\dd s_1}
= \int\limits_{}^{1}\int\limits_{}^{1} \frac{\dd x_1 \dd x_2}
{|1-\Pi(sx_1x_2)|^2}
\frac{\dd\tilde\sigma_B^{j}(p_+x_2,p_-x_1;k_1,q_+,q_-)}
{\dd x_-\dd c\;\dd s_1}
\\ \nonumber && \qquad \times
D(x_1,l_s)D(x_2,l_s)\biggl(1+\frac{\alpha}{\pi}K^{j}\biggr) +
\frac{\alpha}{2\pi}\int\limits_\Delta^1\dd x
\biggl[\frac{1+(1-x)^2}{x}\ln\frac{\theta_0^2}{4}+x\biggr]
\\ \nonumber && \qquad \times
\biggl[\frac{\dd\tilde\sigma^{j}_B(p_-(1-x),p_+;k_1,q_+,q_-)}
{\dd x_-\dd c\;\dd s_1}
+\frac{\dd\tilde\sigma^j_B(p_-,p_+(1-x);k_1,q_+,q_-)}
{\dd x_-\dd c\;\dd s_1}\biggr]
+\frac{\dd\sigma_{\gamma\gamma,\, noncoll}^{j}}{\dd x_-\dd c\;\dd s_1},
\\ \nonumber &&
D(x,l_s)=\delta(1-x)+\frac{\alpha}{2\pi}P^{(1)}(x)(l_s-1)+....,
\; \Delta=\frac{\Delta\varepsilon}{\varepsilon},\; 
P^{(1)}(x)=\biggl(\frac{1+x^2}{1-x}\biggr)_+,\; j=\mu,\pi.
\ea
The boosted cross sections $\dd\tilde\sigma$ is defined above
in Eq.~(\ref{boosted}). The lower limits of the integrals over $x_{1,2}$
depend on the experimental conditions.

The structure function $D$ include all dependence on the large
logarithm $l_s$. The so--called $K$-factor reads
\begin{gather}
K^j=\frac{1}{R^j}B^{\lambda\sigma}\left(i_{\lambda\sigma}^{(vj)}+
i_{\lambda\sigma}^{(sj)}+i_{\lambda\sigma}^{(hj)}\right)
+R^{(j)}_{compt}.
\end{gather}

Quantities $R^{(j)}_{compt}$ include the "non-leading" contributions from
the initial state radiation. Generally, they are rather cumbersome
expressions for the case $s_1\sim s$. For the case $s_1\sim M^2\ll
s$ we obtain
\begin{eqnarray}
R^{(\mu)}_{compt} &=& R^{(\pi)}_{compt}+ \frac{c^2}{(1-2x_-x_++2\sigma)(1+c^2)},
\\ \nonumber
R^{(\pi)}_{compt} &=& \frac{1-c^2}{4(1+c^2)} \bigg\{ \frac{5+2c+c^2}{1-c^2}
\ln^2\bigg(\frac{2}{1+c}\bigg)
-\frac{5-c}{1+c}\ln\bigg(\frac{2}{1+c}\bigg)
\\
&+&\frac{5-2c+c^2}{1-c^2}
\ln^2\bigg(\frac{2}{1-c}\bigg)
-\frac{5+c}{1-c}\ln\bigg(\frac{2}{1-c}\bigg)
-4\frac{c^2}{1-c^2} \bigg\}+\frac{\pi^2}{3}.
\end{eqnarray}
Here we see the remarkable phenomena: the cancellation of terms
containing $\ln(\frac{s}{s_1})$. In such a way only one kind of
large logarithm $\ln(s/m^2)$ enters in the final result. This fact is  
the consequence of the renormalization group invariance. 

\begin{figure} 
\includegraphics[scale=0.8]{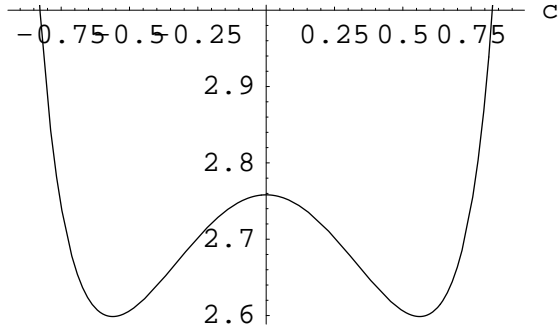}
\caption{ RC  $R^{(\mu)}_{compt}$ for $\sigma=0.1$, $x_-=0.1$}
\label{fig1}
\end{figure}

\begin{figure} 
\includegraphics[scale=0.8]{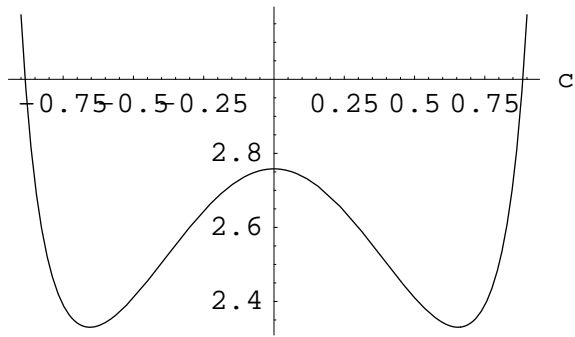}
\caption{ RC   $R^{(\pi)}_{compt}$ for $\sigma=0.1$, $x_-=0.1$}
\label{fig2}
\end{figure}

The values of $R^{(\mu)}_{compt}$, $R^{(\pi)}_{compt}$ are
depicted on fig.\ref{fig1},\ref{fig2}.

\section{Conclusions}

We have considered radiative muon (pion) pair production in high energy
electron-positron annihilation for the charge-blind experimental set--up.
Anyway, the charge--odd part of the cross section under consideration 
is suppressed by the factor $s_1/s\ll 1$ in the kinematics discussed here.

Using the heavy photon Compton tensor \cite{19}, we have calculated 
radiative corrections to this process. Although the analogous calculations 
were performed earlier (see, for example, \cite{11,12a,12b,13}), our main
result, equation (\ref{all}), is new, we believe. This result shows that  
the cross section in our quasi $2\to2$ kinematics can be
written in the form of the cross section of the Drell--Yan process. 
Thus the results of \cite{20}
for the RC to the one-photon $e^+e^-$-annihilation into hadrons are 
generalized to the situation when a hard photon at a large angle is present
in the final state. 

This generalization is not a trivial fact because the two types of 
{\em\ large logarithms} are present in the problem. 

Possible background from the peripherical
process $e\bar{e}\to e\bar{e} \mu\bar{\mu}$ is negligible in our
kinematics: it is suppressed by the factor
$\frac{\alpha}{\pi}\frac{s_1}{s}$ and, besides, can be eliminated if
the registration of the primary hard photon (see Eq.~(\ref{process}))
is required by the experimental cuts for event selection.

\begin{acknowledgments}
One of us (E.A.K.) is grateful to Slovak Institute for Physics,
Bratislava for hospitality when the final part of this work was
done. This work was supported by RFBR grant 03-02-17077 and INTAS
grant 00366.
\end{acknowledgments}

\section*{Appendix}
\setcounter{equation}{0}
\renewcommand{\theequation}{A.\arabic{equation}}

The one--loop QED form factors of muon and pion are
\begin{eqnarray}
&& \mathrm{Re}\,F^{(\mu)}_1(s_1) = 1
+ \frac{\alpha}{\pi}f_1^{(\mu)}(s_1),\quad
\mathrm{Re}\,F^{(\mu)}_2(s_1)=\frac{\alpha}{\pi}f_2^{(\mu)}(s_1)
\nonumber \\
&& f_1^{(\mu)}(s_1)= \biggl(\ln\frac{M}{\lambda} -
1\biggr) \biggl(1 - \frac{1+\beta^2}{2\beta}l_\beta\biggr) +
\frac{1+\beta^2}{2\beta}\biggl( - \frac{1}{4}l^2_\beta +
l_\beta\ln\frac{1+\beta}{2\beta} + \frac{\pi^2}{3}
\nonumber \\
&& \qquad + 2\Li{2}{\frac{1-\beta}{1+\beta}}
\biggr) - \frac{1}{4\beta}l_\beta ,
\nonumber \\
&& f_2^{(\mu)}(s_1)=-\frac{1-\beta^2}{8\beta}l_\beta,
\nonumber \\
&& \mathrm{Re}\,F^{QED}_{\pi}(s_1) = 1 +
\frac{\alpha}{\pi}f^{QED}_{\pi}(s_1),
\nonumber \\
&& f^{QED}_{\pi}(s_1)=\biggl(\ln\frac{M}{\lambda} - 1\biggr) \biggl(1 -
\frac{1+\beta^2}{2\beta}l_\beta\biggr) +
\frac{1+\beta^2}{2\beta}\biggl( - \frac{1}{4}l^2_\beta +
l_\beta\ln\frac{1+\beta}{2\beta} + \frac{\pi^2}{3}
\nonumber \\ && \qquad
+ 2\Li{2}{\frac{1-\beta}{1+\beta}} \biggr), \qquad
\beta^2 = 1 - \frac{4M^2}{s_1},\qquad
l_\beta=\ln\frac{1+\beta}{1-\beta}\, .
\end{eqnarray}

The expressions for leptonic and
hadronic \cite{22} contributions into the vacuum
polarization operator $\Pi(s)$ are:
\begin{eqnarray}
\Pi(s) &=& \Pi_l(s) + \Pi_{h}(s),
\nonumber \\
\Pi_l(s) &=& \frac{\alpha}{\pi}\Pi_1(s)
+ \left(\frac{\alpha}{\pi}\right)^2\Pi_2(s)
+ \left(\frac{\alpha}{\pi}\right)^3\Pi_3(s) + \dots
\nonumber \\
\Pi_h(s) &=& \frac{s}{4\pi^2\alpha}\Biggl[
\mathrm{PV}\int\limits_{4m_{\pi}^2}^{\infty}
\frac{\sigma^{e^+e^-\to\mathrm{hadrons}}(s')}{s-s'}\dd s'
- \mathrm{i}\pi\sigma^{e^+e^-\to\mathrm{hadrons}}(s)\Biggr].
\end{eqnarray}
The first order leptonic contribution is well known \cite{23}:
\begin{equation}
\Pi_1(s) = \frac{1}{3}\ln\frac{s}{m^2} - \frac{5}{9} + f(x_{\mu}) + f(x_{\tau})
- \mathrm{i}\pi\left[\frac{1}{3} + \phi(x_{\mu})\Theta(1-x_{\mu})
+ \phi(x_{\tau})\Theta(1-x_{\tau})\right],
\end{equation}
where
\begin{eqnarray*}
f(x)\!&=&\! \left\{\begin{array}{l}
\!\!-\frac{5}{9}-\frac{x}{3}+\frac{1}{6}(2+x)\sqrt{1-x}\;\ln\left|
\frac{1+\sqrt{1-x}}{1-\sqrt{1-x}}\right|\ \ \ \,\mathrm{for}\ \ x\leq 1, \\
\!\!-\frac{5}{9}-\frac{x}{3}+\frac{1}{6}(2+x)\sqrt{x-1}\;\arctan\left(
\frac{1}{\sqrt{x-1}}\right)\ \ \ \mathrm{for}\ \  x > 1,\\
\end{array}\right. \\
\phi(x)\!&=&\! \frac{1}{6}(2+x)\sqrt{1-x},\qquad
x_{\mu,\tau} = \frac{4m_{\mu,\tau}^2}{s}\, .
\end{eqnarray*}
In the second order it is enough to take only the logarithmic term
from the electron contribution \cite{24}
\begin{equation}
\Pi_2(s) = \frac{1}{4}(\ln\frac{s}{m^2}-\mathrm{i}\pi) + \zeta(3) - \frac{5}{24}\, .
\end{equation}


\begin{thebibliography}{99}
\bibitem{1}
M.~S.~Chen and P.~M.~Zerwas,
Phys.\ Rev.\ D {\bf 11}, 58 (1975).
\bibitem{2}
M.~W.~Krasny, W.~Placzek and H.~Spiesberger,
Z.\ Phys.\ C {\bf 53}, 687 (1992).
\bibitem{3}
A.~B.~Arbuzov, E.~A.~Kuraev, N.~P.~Merenkov and L.~Trentadue,
JHEP {\bf 9812}, 009 (1998).
\bibitem{3a}
H.~Anlauf, A.~B.~Arbuzov, E.~A.~Kuraev and N.~P.~Merenkov,
JHEP {\bf 9810}, 013 (1998);
Phys.\ Rev.\ D {\bf 59}, 014003 (1999).
\bibitem{4}
S.~Spagnolo,
Eur.\ Phys.\ J.\ C {\bf 6}, 637 (1999).
\bibitem{5}
S.~Binner, J.~H.~Kuhn and K.~Melnikov,
Phys.\ Lett.\ B {\bf 459}, 279 (1999).
\bibitem{6}
M.~Benayoun, S.~I.~Eidelman, V.~N.~Ivanchenko and Z.~K.~Silagadze,
Mod.\ Phys.\ Lett.\ A {\bf 14}, 2605 (1999).
\bibitem{7}
X.~C.~Lou, T.~Benninger and W.~M.~Dunwoodie,
Nucl.\ Phys.\ A {\bf 675}, 253C (2000).
\bibitem{8}
A.~G.~Denig {\it et al.}  [the KLOE Collaboration],
Nucl.\ Phys.\ Proc.\ Suppl.\  {\bf 116}, 243 (2003).
\bibitem{9}
E.~P.~Solodov  [BABAR collaboration],
in {\it Proc. of the $e^+ e^-$ Physics at Intermediate Energies Conference } 
ed. Diego Bettoni, eConf {\bf C010430}, T03 (2001) [hep-ex/0107027].
\bibitem{10}
A.~B.~Arbuzov et al.,
JHEP {\bf 9710}, 006 (1997).
\bibitem{11}
A.~Hoefer, J.~Gluza and F.~Jegerlehner,
Eur.\ Phys.\ J.\ C {\bf 24}, 51 (2002).
\bibitem{12}
H.~Czyz, A.~Grzelinska, J.~H.~Kuhn and G.~Rodrigo,
Eur.\ Phys.\ J.\ C {\bf 27}, 563 (2003);
H.~Czyz, J.~H.~Kuhn and G.~Rodrigo,
Nucl.\ Phys.\ Proc.\ Suppl.\  {\bf 116}, 249 (2003);
G.~Rodrigo, H.~Czyz and J.~H.~Kuhn,
eConf {\bf C0209101}, WE06 (2002) [hep-ph/0210287].
\bibitem{12a}
G.~Rodrigo, H.~Czyz, J.~H.~Kuhn and M.~Szopa,
Eur.\ Phys.\ J.\ C {\bf 24}, 71 (2002).
\bibitem{12b}
J.~H.~Kuhn and G.~Rodrigo,
Eur.\ Phys.\ J.\ C {\bf 25}, 215 (2002).
\bibitem{13}
V.~A.~Khoze et al.,
Eur.\ Phys.\ J.\ C {\bf 18}, 481 (2001);
Eur.\ Phys.\ J.\ C {\bf 25}, 199 (2002).
\bibitem{14}
J.~Gluza, A.~Hoefer, S.~Jadach and F.~Jegerlehner,
Eur.\ Phys.\ J.\ C {\bf 28}, 261 (2003).
\bibitem{15}
F.~A.~Berends, R.~Kleiss, S.~Jadach and Z.~Was,
Acta Phys.\ Polon.\ B {\bf 14}, 413 (1983).
\bibitem{16}
A.~B.~Arbuzov et al.,
JHEP {\bf 9710}, 001 (1997).
\bibitem{17}
F.~A.~Berends and R.~Kleiss,
Nucl.\ Phys.\ B {\bf 177}, 237 (1981);
E.~A.~Kuraev and G.~V.~Meledin,
Nucl.\ Phys.\ B {\bf 122}, 485 (1977).
\bibitem{18}
A.~B.~Arbuzov et al.,
JHEP {\bf 9710}, 006 (1997).
\bibitem{19}
E.~A.~Kuraev, N.~P.~Merenkov and V.~S.~Fadin,
Yad.\ Fiz.\  {\bf 45}, 782 (1987).
\bibitem{Baier}
P.~Kessler, 
Nuovo Cim.\  {\bf 17}, 809 (1960);
V.~N.~Baier, V.~S.~Fadin and V.~A.~Khoze,
Nucl.\ Phys.\ B {\bf 65}, 381 (1973);
V.~N.~Baier, E.~A.~Kuraev, V.~S.~Fadin and V.~A.~Khoze,
Phys.\ Rept.\  {\bf 78}, 293 (1981).
\bibitem{20}
E.~A.~Kuraev and V.~S.~Fadin,
Sov.\ J.\ Nucl.\ Phys.\  {\bf 41}, 466 (1985);
O.~Nicrosini and L.~Trentadue,
Phys.\ Lett.\ B {\bf 196}, 551 (1987).
\bibitem{22}
S.~Eidelman and F.~Jegerlehner,
Z.\ Phys.\ C {\bf 67}, 585 (1995).
\bibitem{23}
A.~I.~Akhiezer and V.~B.~Berestetski, Quantum Electrodynamics, 1981.
\bibitem{24}
C.~R.~Hagen and M.~A.~Samuel,
Phys.\ Rev.\ Lett.\  {\bf 20}, 1405 (1968).
\bibitem{viol}
A. Arbuzov et al, JETP (88) 1999, p.213
\end{thebibliography}
\end{document}